\documentclass{ws-p9-75x6-50}
\input psfig.tex
\newcommand{\gsim}{\,\raisebox{-0.4ex}{$\stackrel{>}{\scriptstyle\sim}$}\,}

\begin{document}

\title{Understanding Galactic Black Hole Candidate\\
 GRS 1915+105}

\author{Sandip K. Chakrabarti, S.G. Manickam and A. Nandi}

\address{S.N. Bose National Centre for Basic Sciences, JD-Block, Salt Lake, Calcutta
700098\\E-mail: chakraba@boson.bose.res.in, sivman@boson.bose.res.in and anuj@boson.bose.res.in}

\author{A.R. Rao}

\address{Tata Institute of Fundamental Research, Homi Bhabha Road, Colaba, 400005
\\E-mail: arrao@tifr.res.in}

\maketitle

\abstracts{We interpret observed features of the black hole candidate GRS1915+105 in 
the light of the advective flow paradigm. The light curve of this black hole
is seen to be of four classes, Soft, Hard, Semi-Soft and Intermediate. 
Each class can display more than one type of light curve making altogether twelve or more
types of light curves. In each type, the mode of accretion may be one or more of three types:
which we identify as HW (hard state with winds), CC (Compton cooling) and EA (enhanced accretion) 
respectively. Apart from these three states there are two more fundamental states
in a black hole accretion which are soft and hard states. We also find that winds and outflows could
soften spectra in hard states and the enhanced accretion could harden the soft states. These are observed in 
the spectra of GRS 1915+105. Generally, we find that advective flow paradigm can satisfactorily explain
all the features of the enigmatic black hole GRS 1915+105.} 

\noindent Proceedings of the 9th Marcel Grossman conference Ed. R. Ruffini

\section{Introduction}

In the standard model of a thin disk, matter moves in a Keplerian orbit since pressure 
gradient force and inertial force were ignored. Furthermore, the flow is truncated at the
last stable circular orbit at $r=6GM/c^2$. When these forces are included, matter is seen to deviate
from a Keplerian flow last few tens to hundreds of Schwarzschild radii ($r_g=2GM/c^2$).
The flow must pass through the horizon supersonically, and therefore must acquire a 
considerable radial motion in the last few Schwarzschild radii. Infall time becomes 
very small compared to the timescale of transport of angular momentum, and matter
moves with almost constant angular momentum which results in producing centrifugal barrier
dominated boundary layer or CENBOL. CENBOL could be due to a sharp steady shock or just a region with
gradual rise in density. The basic solutions of the accretion and winds are presented in
several places\cite{chak96,cimen00,mg9} and will not be discussed here.
For easy reference we direct to Figs. 1(A-J) and Figs. 1(a-j) of Chakrabarti\cite{mg9} which describe
the building blocks of the  black hole accretion.
When these basic building blocks are combined, very realistic accretion and wind flow models are
obtained\cite{samar00,cimen00}. Figs. 1(a-e) show five distinct combinations\cite{chaknan00} of accretion and wind flows
which we term as the fundamental states. Typically, the flow is assumed to have both the
Keplerian component ${\dot M}_d$ and the sub-Keplerian component ${\dot M}_h$.
When Keplerian component is
very low ($\sim 0.001-0.1 {\dot M}_{Edd}$) but sub-Keplerian component is high (${\dot M}_d \sim 1{\dot M}_{Edd}$), 
soft photons are fewer in number and electrons remain hot and emit very hard X-rays. 
Shocks may not be produced but the inner edge is puffed up due to the centrifugal force (Fig. 1a).
This is the canonical hard state (H) which may or may not display 
Quasi-Periodic Oscillations (QPO). Very little winds are produced. In the Off-state (HW),
accretion rates could be as above but the shocks may form. This is possible when viscosity is 
lower than the critical value and the net energy of the Keplerian and sub-Keplerian matter is positive. 
Outflows are produced\cite{chak99} at a very low but steady rate (Fig. 1b).  As the sonic sphere 
(subsonic region below the sonic point of the outflow) is gradually filled up, the radiation 
gets softer. QPOs may or may not be observed depending on whether shocks oscillate\cite{msc96,rcm97} or not.
In the `Dip State' (CC), Keplerian rates are higher (${\dot M}_K \sim 0.1-0.3 {\dot M}_{Edd}$) and 
viscosity is also higher so that shocks are weaker. Outflow till the sonic 
sphere achieves sufficient optical depth that it is cooled down by Comptonization. 
The sonic point comes closer to the black hole as sound speed goes down in this region. 
Flow which remains sub-sonic with respect to this sonic sphere losses outward drive 
and returns back to the disk, while the supersonic flow separates as blobs in the 
jets (Fig. 1c). The energy spectral slope is larger compared to that in the hard state. 
QPO may not be visible as the post-shock region is cooler (with a very large cooling 
time scale) while infall time is shorter since the Keplerian disk moves inward due to 
larger turbulent and radiative viscosity. In the so called high count or `On State' (EA), 
the flow parameters may remain similar to above, but the return flow enhances 
both Keplerian and sub-Keplerian disk rates {\it locally} in the last few hundred 
Schwarzschild radii (Fig. 1d).  Duration of this state is the duration of drainage 
of the excess accretion (EA) from return flow and the viscosity becomes high enough 
(more than critical) so that Keplerian disk moves in all the way to marginally stable orbit.
Softer state with a high photon flux  should be seen and no QPO should be observed. In case,
enhanced accretion is not very high, outflow remains active and the accretion may {\it perpetually}
remain in State CC. For the transition from CC to EA, the accretion must rise sufficiently and
most of the matter must accretion rather than flying away as winds.
Finally, in the `Soft State' (S), accretion rate of the Keplerian component is very high
(${\dot M}_K\gsim 0.3 {\dot M}_{Edd}$). Here, CENBOL is totally cooled, shock is totally disappeared, 
outflow is cut-off and 
matter moves almost radially (Fig. 1e) and transfers its momentum to soft photons by Bulk 
Motion Comptonization\cite{ct95}. The spectrum is softer with a weak power-law 
hard-tail and no QPO should be seen. In Fig. 1f we identify suggestive regions in the
$R_{\dot m}$ = outflow rate/inflow rate vs. compression ratio $R$ space with the fundamental states
given above. Hard states without shock do not fall in this diagram.

Compression   It is to be noted that there could be states intermediate to 
these states, since the mass flux can vary continuously. However
transitions from HW (C) to CC (A), may be quick, since it involves rapid Compton cooling.

Chakrabarti \& Titarchuk\cite{ct95} and Chakrabarti\cite{chak97} discussed 
mainly hard and soft states. In several occasions variation of these 
two states are also discussed\cite{samar00,cetal00,cimen00}
Chakrabarti and Nandi\cite{chaknan00} systematically identified these
states to five distinct types to encompass more complex situations involving winds and radiative transfer.
The intriguing black hole candidate GRS1915+105 seems to definitely require winds
to describe its lightcurves and other time dependent behaviours. In the rest of this paper we try to understand
this object in the light of the Chakrabarti-Nandi state descriptions as above.

\begin{figure}
\vbox{
\vskip 0.0cm
\hskip 0.0cm
\centerline{
\psfig{figure=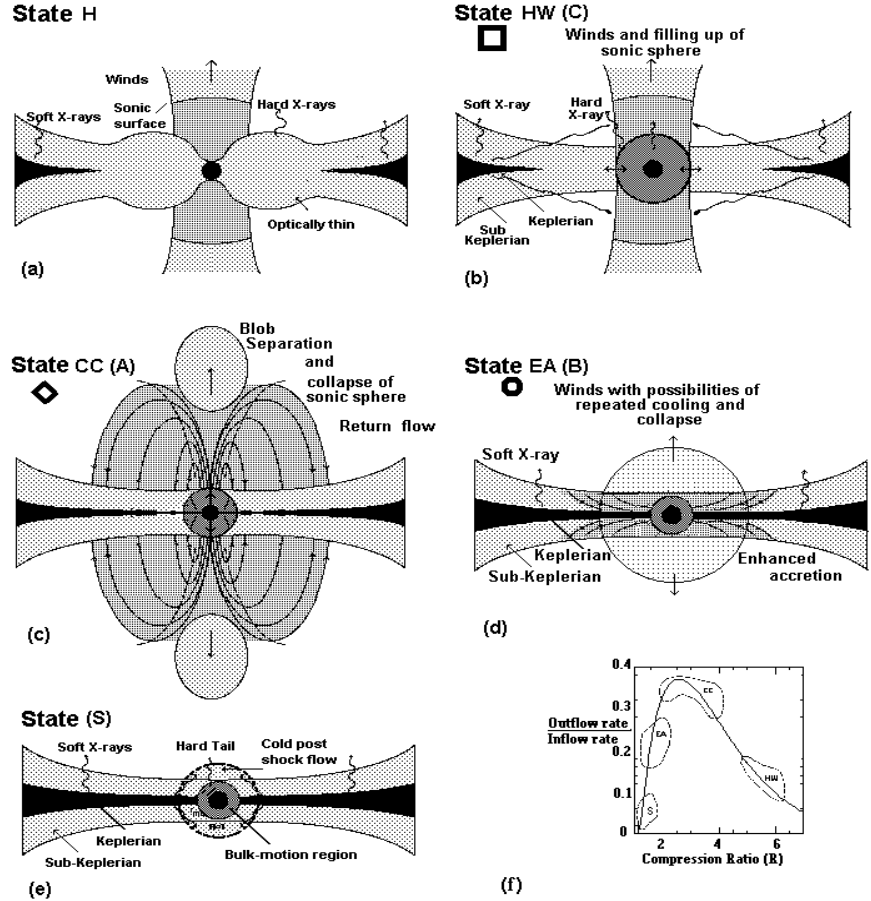,height=12truecm,width=12truecm}}}
\vspace{0.0cm}
\caption[]{Schematic diagram of the flows in five possible fundamental modes of accretion (a-e). GRS1915+105
stays most of the time in State HW, and moves to CC and EA to form various classes of light curves. 
In parenthesis, we mark A,B,C to identify conjectured States of Belloni et al.\cite{belo00}. In (f) we 
suggest where these states fall in a $R_{\dot m}$ vs. $R$ diagram.}
\end{figure}

\section{Light curves of GRS 1915+105}

\subsection{Classification}

Fig. 2 shows all possible types of light curves\cite{nanchak}. These are placed in
twelve panels which were designated as $\chi$, $\alpha$, $\nu$, $\beta$, $\lambda$,
$\kappa$, $\rho$, $\mu$, $\theta$, $\delta$, $\gamma$ and $\phi$ respectively by Belloni et al\cite{belo00}. When 
softness ratio diagram is drawn with A/C vs. B/C (where A:0-3keV; B:3-17keV; 
C:17-60keV), they reveal four basic classes\cite{nanchak}:\\
\noindent Panel 1: HARD (H) Class (Panel 1), 2. INTERMEDIATE  (I) Class (Panels 2-8), 3. SEMI-SOFT (SS)
Class (Panel 9) and 4. SOFT (S) Class (Panels 10-12).\\
In each Class, the flow may stay in one or more States of Fig. 1\cite{chaknan00}.
For instance, Belloni et al\cite{belo00} 
showed that in all the light curves the flow stays in three states A, B, and C,
with varying sequence and combination.  We believe that these states
are the same as our CC (Compton cooled wind), EA (Enhanced accretion) and HW (Hard
states with winds) respectively. Belloni et al. also observed that a direct transition from State C to State B 
is prohibited and  the flow must pass through the State A. In view of the above classification of States, we can 
understand the reason. When the outflow rate is very high, the sonic sphere may be cooled down after the
optical length reaches unity (thus reaching CC from HW State). The cooled flow returns back to the disk and 
causes enhanced accretion (EA). EA is not possible without going through the state CC. On the other hand,
after EA, CC can form since outflow rate also goes up and CC State may form with enhanced accretion.
Similarly, after CC, if the entire matter from the sonic sphere separates as blobs, 
flow can come back to HW state. It is to be noted that, our separation of four classes,
which was based on soft-ness diagrams, falls in line with Belloni et al.'s observation
that H class is in C State, I Class is in CABCAB.. States, SS Class is in CACA.. States
and S Class is in ABAB ... States.

\begin{figure}
\vbox{
\vskip -5.0cm
\hskip 0.0cm
\centerline{
\psfig{figure=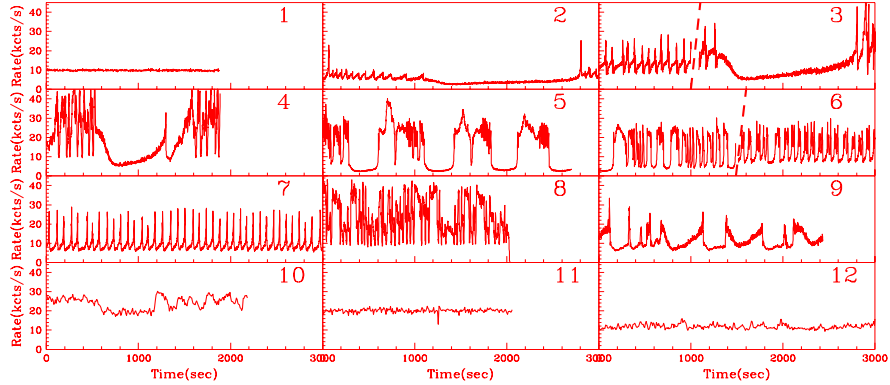,height=15truecm,width=12truecm}}}
\vspace{-5.0cm}
\caption[]{
All possible light curves of the black hole candidate GRS 1915+105. Along X-axis
is time in seconds and along Y-axis is photon counts in units of $10^3$. In Panels 3 and 6
two days of curves, differing slightly, have been put.
}
\end{figure}

Based on the discussions given above, we can present a feedback diagram\cite{chakhd00} in Fig. 3 which shows that
in order to fully understand the problem, the function ${\dot M}_{out} = {\dot M}_{out} ({\dot M}_{in})$
and the disk viscosity which separates net accretion into  Keplerian and sub-Keplerian rates
must be known. The geometry of the CENBOL and sonic sphere which dictates
how many soft photons would be intercepted  [S($\gamma$) in Fig. 3] also plays a crucial role in
bringing the spectrum to softer state without raising the Keplerian accretion rate. Here,
${\dot M}_d$ is the Keplerian rate and ${\dot M}_h$ is the sub-Keplerian rate, ${\dot M}_{rf}$ is the 
amount by which return flow takes place.

\begin{figure}
\vbox{
\vskip -0.0cm
\hskip 0.0cm
\centerline{
\psfig{figure=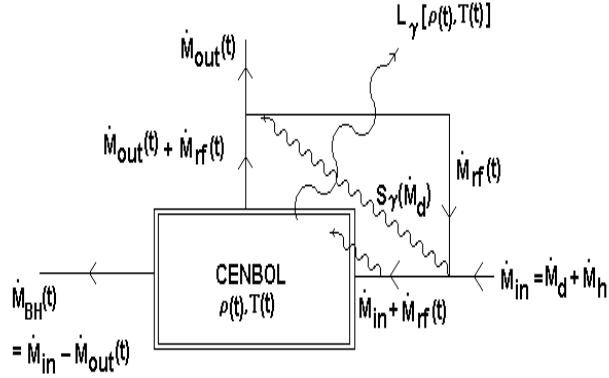,height=10truecm,width=10truecm}}}
\vspace{-3.5cm}
\caption[]{
Schematic diagram showing the factors involved in causing non-linearity in the $L_\gamma $ (t) the light curve.}
\end{figure}

\subsection{Correlation between Duration and Frequency of QPO}

Chakrabarti \& Manickam\cite{cm00} showed that there is a distinct correlation between the
average QPO frequency ($\nu_{QPO}$) and the duration ($t_{off}$) of QPOs 
in light curves where high and low count states (`On' and `Off' states) 
are present. These correspond to I and SS classes\cite{nanchak}. 
Chakrabarti \& Manickam\cite{cm00} argue that $t_{off}\propto \nu^{-2}$s. 
This relation shown in Fig. 4 seems to agree with the observation. 
The basis for this argument is as follows:
Chakrabarti\cite{chak99} showed that for an outflow, which is isothermal 
at least till the sonic sphere, the location of the sonic surface ($R_c$) and the 
shock location ($R_s$) in accretion are related by $R_c \sim f_0 R_s/2$, where,
$f_0=R^2/(R-1)$, with $R$ as the compression ratio of the flow at the shock. For a very strong shock $R\sim 7$
and $R_c \sim 4 R_s$. Because $R_c$ is very far out, sonic sphere is difficult to 
fill in and it may not be cooled by intercepted soft photons from the Keplerian disk. 
This is why the presence of QPOs does not necessarily imply that On and Off transitions.
The ratio of the outflow to inflow rate ${R_{\dot m}}$ shows a peak at around $R\sim 2.5$
where $R_c$ is much closer: $R_c \sim 2 R_s$. Since the time-period of oscillation of shocks 
is proportional to the infall time inside CENBOL\cite{cm00}, and the filling time of the sonic 
sphere is proportional to its volume, both the times are related to the shock location $R_s$.
It can be easily shown that\cite{cm00}, 
$$
t_{off} =461.5(\frac{0.1}{\Theta_{\dot m}}) 
(\frac{M}{10M_\odot})^{-1} (\frac{v_0}{0.066})^2 \nu^{-2},
$$
where, $\Theta_{\dot m}=\frac{\Theta_{out}}{\Theta_{in}} \frac{{\dot M}_{in}}{{\dot M}_{Edd}}$
is a factor directly related to the solid angles subtended by the inflow ($\Theta_{in}$),
outflow ($\Theta_{out}$) and the net (Keplerian plus sub-Keplerian) accretion rate 
${\dot M}_{in}$ in units of the Eddington rate ${\dot M}_{Edd}$. $v_0$ is the average velocity
of the post-shock region in units of the velocity of light. This relation has since then been
found to be valid for all the days of observations.

\begin{figure}
\vbox{
\vskip -3.0cm
\hskip 0.0cm
\centerline{
\psfig{figure=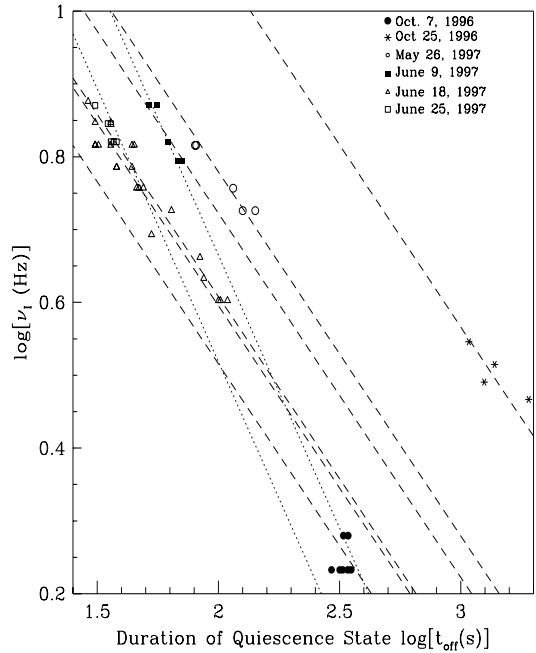,height=12truecm,width=8truecm}}}
\vspace{0.0cm}
\caption[]{
Variation of QPO frequency $\nu$ with duration of off states $t_{off}$. 
Dotted curves are the $t_{off} \propto \nu^{-4/3}$ law derived using simple
free-fall velocity assumption in the post-shock region and dashed curves 
are the $t_{off}\propto \nu^{-2}$ with constant post-shock velocity law\cite{cm00}. }
\end{figure}

A good way to test whether QPOs are generated because of the oscillation of CENBOL is to 
do power density spectral analysis for photons belonging to different energy channels. 
Fig. 5 shows a result\cite{cm00} which is very general and is found to be valid 
for all the days of observations tested so far.  Here, power is plotted  
against the average frequency of QPO for various energy channels marked in each 
panel. Clearly, in low-energies (0-4keV) QPO is absent. QPO is strong only in high energies.
This agrees with the understanding that the pre-CENBOL soft-photons do not participate in
oscillations. This  was tested more stringently in Rao et al\cite{rao00} by actually
separating black body photons and Comptonized power-law photons. They find that only 
Comptonized photons participate in QPOs.

\begin{figure}
\vbox{
\vskip -1.5cm
\hskip 0.0cm
\centerline{
\psfig{figure=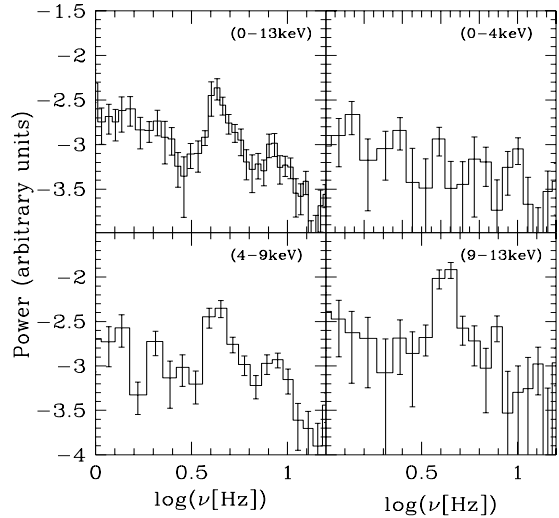,height=12truecm,width=10truecm}}}
\vspace{-3.0cm}
\caption[]{Power density spectrum of an  off-state
constructed from selected channel intervals of the
binned RXTE-PCA data. QPO is seen only in high energies, 
strongly pointing to the shock oscillation model. }
\end{figure}

\subsection{Origin of the Splitting of Peaks}

Occasionally, one observes that at the onset of the high count or 
the On-state, there are two distinct peaks and the second peak is 
harder compared to the first peak\cite{nanetal00}. This is demonstrated 
in Fig. 6 where the light curves are shown (RXTE observation ID: 10408-01-41-00) 
in four panels and the energy channels are marked\cite{nanetal00}. The ratio of 
the photon counts at the two peaks reach almost unity at very high energies while
at low energies the second peak is very weak. Though it is not an universal observation,
the off state and the first peak (P1) is found to have a QPO while the second
peak (P2) does not show QPO. 

We believe that these phenomena could be understood easily using our paradigm\cite{chaknan00}
as shown in  Fig. 1(a-e). In this scenario, P1 is a continuation of the off state
(HW) and the dip in between the P1 and P2 is due to Compton cooling (CC). P2 (or similar noisy peaks
on wide on states, for that matter) is due to enhanced accretion (EA). Very often 
we note that when many noisy peaks are observed in broad `on' states, the valleys have QPOs
and the peaks do not. We believe that these valleys are made up of mini-$\rho$
or $\nu$ classes\cite{belo00}.

\begin{figure}
\vbox{
\vskip -1.5cm
\hskip 0.0cm
\centerline{
\psfig{figure=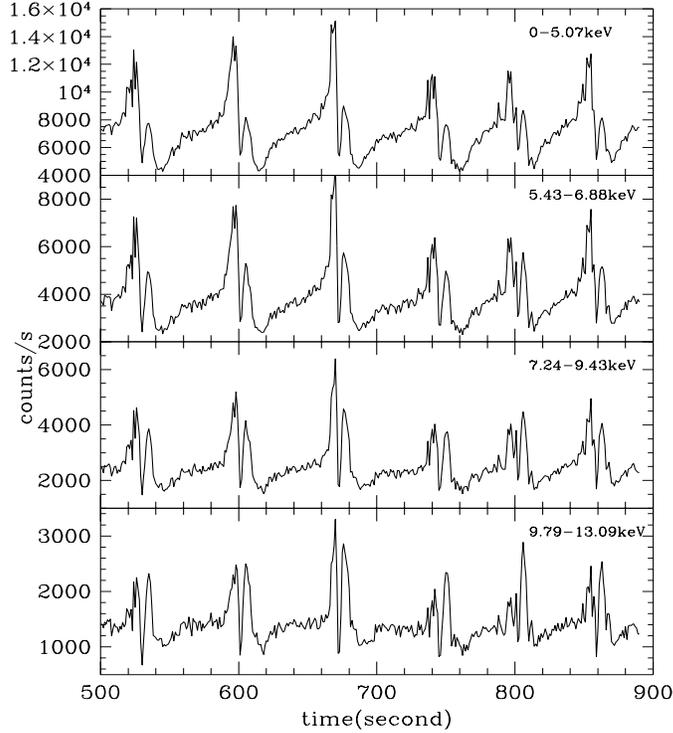,height=12truecm,width=10truecm}}}
\vspace{0.0cm}
\caption[]{
Light curves of a 
part of the observation on Oct. 15th, 1996 (RXTE observation ID: 10408-01-41-00) 
at different channels (energies are marked). Note that ratio of photon counts 
in the two peaks tends to become unity at higher energies. }
\end{figure}

\section{Evidence for Winds in the Spectra}

A number of observations  have hinted, after finding distinct correlation between IR, Radio
and X-rays, that winds must be produced from regions very close to the black hole (e.g., 
Dhawan et al.\cite{d00}), vindicating the advective paradigm model that both 
the Comptonized radiation and the outflowing winds originate from the CENBOL. 
We present here a further, and a more direct evidence that 
winds are indeed produced from the CENBOL without taking resort to IR and Radio measurements.
Chakrabarti\cite{chak98} pointed out that if winds are formed out of the CENBOL, then the spectrum
should be softened, since there are fewer electrons to cool by the same amount of soft-photon flux
from the pre-CENBOL flow. A corollary is that when cool matter from  the sonic sphere
returns back to the CENBOL, the spectrum should be hardened. A bye-product of these two effects 
(hardening of soft states and softening of hard states)
is that the pivotal point where the power-law hard tails of these two states intersect, should be
shifted to a much higher energy. In Fig. 7, we present the theoretical prediction
based on the computation of spectra with (solid) or without (dotted) taking effects of winds.
Disk accretion rates for the soft and the hard states are chosen to be $0.3 {\dot M}_{Edd}$
and $0.1 {\dot M}_{Edd}$ respectively. Sub-Keplerian halo accretion rate is kept fixed at
$1.0 {\dot M}_{Edd}$ and the shock is located at $R_{s}=14$ and $R_s=10$
respectively in low and high states. This is in line
with the general conclusion that the inner edge moves in during soft states.
Other parameters are kept identical. The solid curves are the
off and the on state spectra. In these cases the disk accretion rates are kept
as before, but the twenty percent of CENBOL matter is assumed to be lost
in wind in the off state and ten percent of matter is assumed to be
falling back on the halo from the wind region in the on state.
Because of softening and hardening effects discussed above, the intersection point
is located at a much higher energy.

\begin {figure}
\vbox{
\vskip -6.0cm
\hskip 0.0cm
\centerline{
\psfig{figure=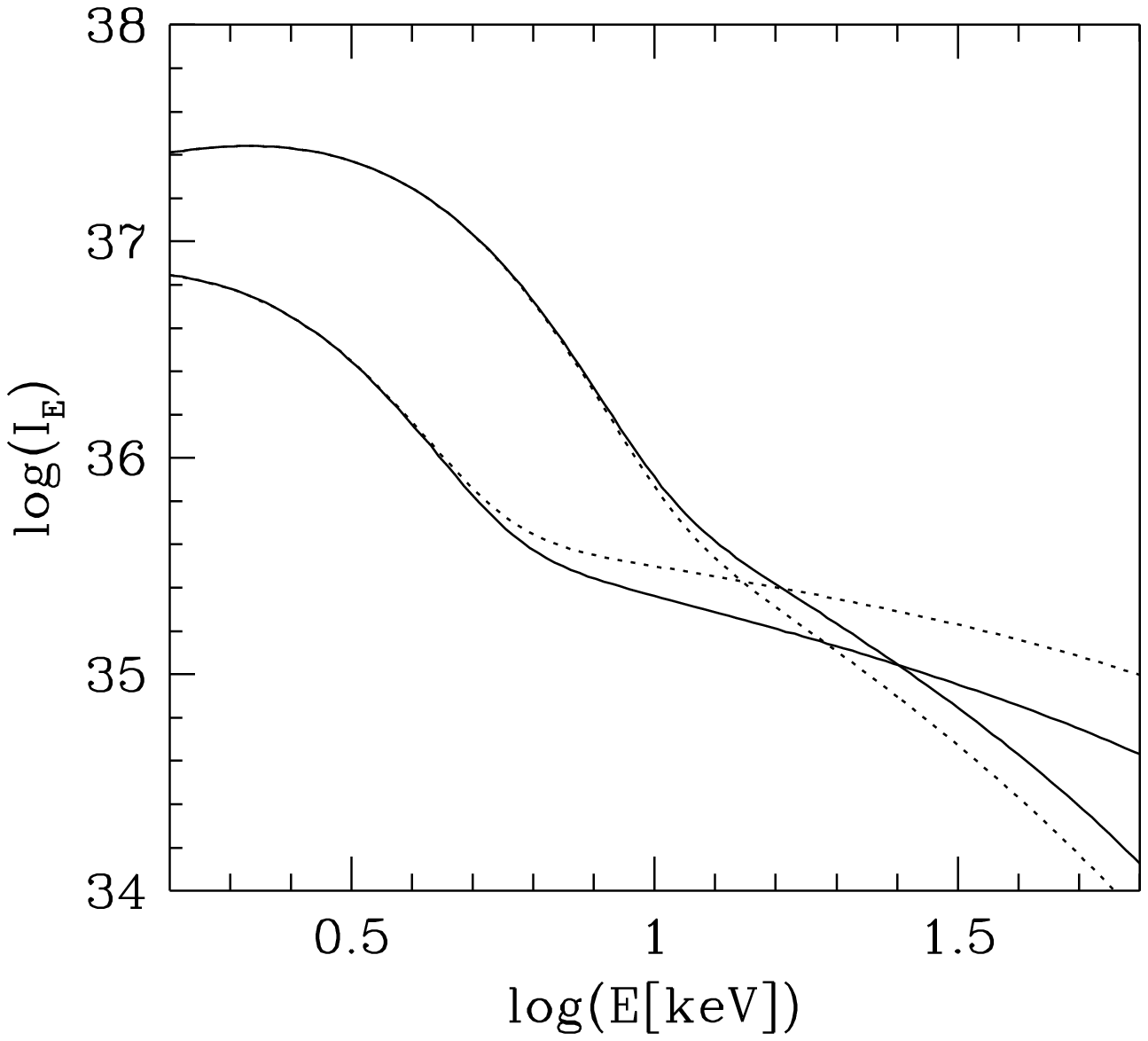,height=15truecm,width=10truecm}}}
\end{figure}
\begin{figure}
\vspace{-3.0cm}
\caption[]
{Typical calculated spectra (uncorrected for absorption)
for off and on states and low and high states of GRS1915+105.
In off state, spectrum appears to be softened with respect to low state, and in on state,
spectrum appear to be hardened with respect to the high state. }
\end{figure}

The effects described above are routinely observed in GRS1915+105\cite{cetal00} 
whenever transitions between high-count and low-count take place. 
Fig. 8 presents\cite{cetal00,chakhd00} RXTE spectra of three separate days of
observations and compares them with the low and high state
spectra of 1997 March 26 (PID No. 20402-01-21-00)  and of 1997 August 19 with
(PID No. 20402-01-41-00)\cite{muno99} respectively. The PID of the three RXTE
observations are (a) 1997 June 18 (PID 20402-01-33-00), (b) 1997 July 10 (PID 20402-01-36-00)
and (c) 1997 July 12 (PID 20402-01-37-01) respectively. In all these days,
the intersections or the pivotal points generated by the `on' and `off' state spectra
are farther out compared to the the pivotal points created by pure
low and high states. This is a clear evidence that significant winds are present in
off states (State HW in Fig. 1) and a significant return flow followed by an 
enhanced accretion is present in the on states (State EA in Fig. 1).
\begin {figure}
\vbox{
\vskip -9.5cm
\hskip 0.0cm
\centerline{
\psfig{figure=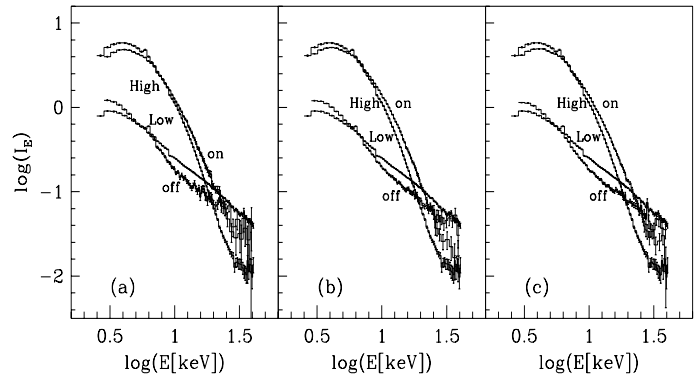,height=15truecm,width=10truecm}}}
\vspace{0.0cm}
\caption[]{Unfolded RXTE-PCA spectra of GRS~1915+105 obtained during the low and high states
are compared with high-count states (On-states) and low-count states (Off-states)
spectra during the irregular bursts observed on (a) 1997 June 18,
(b) 1997 July 10, and (c) 1997 July 12. Histograms show fitted models.
Because of softening of hard states and hardening of soft states, the pivoting occurs
at a higher energy (Figure taken from Chakrabarti et al\cite{cetal00}).}
\end{figure}

\section{Concluding Remarks}

We established that accretion process in a black 
hole could be very complex and for a true understanding
of the spectra and the light curves, one requires 
both the Keplerian and sub-Keplerian flows, centrifugal pressure
dominated boundary layer (CENBOL), winds from CENBOL
and effects of radiative transfer on the sonic sphere and the CENBOL. Most of 
crucial observations could be qualitatively explained in this way. We have also presented
a few fundamental states of accretion and identified three of them with the three states
conjectured by Belloni et al.\cite{belo00}. Minor variation of these basic understanding
cannot be ruled out.

\section{Acknowledgments}

This project is partly supported by a DST grant No. SP/S2/K-14/98 and partly by ISRO RESPOND 
Project `Quasi-Periodic Oscillations of Black Holes'.

\end{document}